\documentclass[twocolumn,pre,superscriptaddress,epfs,aps,showpacs]{revtex4}
\usepackage{amsmath}
\usepackage{graphicx}
\usepackage{color}
\usepackage{epsfig}
\usepackage{braket}	
\usepackage{latexsym}
\usepackage{array} 
\usepackage{multirow}

\begin{document}

\title{Steady states of non-axial dipolar rods driven by rotating fields}
\author{Jorge L. C. Domingos}
\affiliation{Departamento de
F\'{\i}sica, Universidade Federal do Cear\'a,  Caixa Postal 6030, 60455-760 Fortaleza, Cear\'a, Brazil}
\author{Everton A. de Freitas}
\affiliation{Departamento de
F\'{\i}sica, Universidade Federal do Cear\'a,  Caixa Postal 6030, 60455-760 Fortaleza, Cear\'a, Brazil}
\author{W.P. Ferreira}
\affiliation{Departamento de
F\'{\i}sica, Universidade Federal do Cear\'a,  Caixa Postal 6030, 60455-760 Fortaleza, Cear\'a, Brazil}

\date{\today}

\begin{abstract}
We investigate a two-dimensional system of magnetic colloids with anisotropic geometry (rods) subjected to an oscillating external magnetic field. The structural and dynamical properties of the steady states are analyzed, by means of  Langevin Dynamics simulations, as a function of the misalignment of the intrinsic magnetic dipole moment of the rods with respect to their axial direction, and also in terms of the strength and rotation frequency of an external magnetic field. The misalignment of the dipole relative to their axial direction is inspired by recent studies, and this is extremely relevant in the microscopic aggregation states of the system. The dynamical response of the magnetic rods to the external magnetic field is strongly affected by such a misalignment. Concerning the synchronization between the magnetic rods and the direction of the external magnetic field, we define three distinct regimes of synchronization. A set of steady states diagrams are presented, showing the magnitude and rotation frequency intervals in which the distinct self-organized structures are observed.
\end{abstract}


\maketitle

\section{Introduction}
Anisotropic colloids are an exciting and relevant system due to their primary role in soft matter systems. The possibility of manipulation of their shapes and interaction brings a wide range of applications in the self-assembly of colloidal matter \cite{CDias,Rutkowski,quadrupolar}, in microfluidics \cite{Nisisako,Daddi,Debnath} and in the design of functional devices such as probes and sensors \cite{Adrian,Mehdi,Savka,Howes,cousin}. As a particular class of colloids with anisotropic interaction, we mention magnetic colloids, which are micro-sized building blocks with an embedded magnetic moment with dipolar interaction, i.e., particles composed of a magnetic mono-domain having a typical size ranging from $1$ to $150$ $nm$  \cite{Gubin}. 

Magnetic nanoparticles (MN) with anisotropic shape are subject to higher interest in comparison to their spherical counterparts due to their more complex properties and collective behavior, such as magnetic birefringence \cite{lemaire2002ferr}, thermal conductivity \cite{philip2007enhancement} and orientation ordering transition \cite{jorgedomingos,alvarez}. Attention was also addressed to cases where the anisotropy is in the location of the dipole concerning the center of symmetry of the particle. In recent theoretical works, the structure of fluids containing spherical particles with embedded off-centered magnetic dipoles \cite{abrikosov2013self,yener2016self} was investigated.

Concerning the interaction between rod-shaped MN, the magnetic moment is usually parallel to their long axis. However, anisotropic shaped particles with non-axial dipole moment is being subject of many studies. It is known that ellipsoidal particles with permanent dipole moment perpendicular to their long axis showed to be useful in cell entrapment by magnetic manipulation \cite{tierno}. Other studies analyzed rods with non-axial dipole moment e.g. experimentally through superparamagnetic magnetoresponsive rods \cite{singh2005rigid} and peanut-shaped particles \cite{small}, also numerically \cite{jdomingos2} and analytically \cite{carrey}. The latter studied the torque of magnetic rods with randomly oriented dipoles for optimized hyperthermia applications. Therefore, in this study, we explore the direction of the dipole moment as a controlling parameter, allowing the functionalization of the particles.

In this work, we use a peapod-like model to simulate the rigid magnetic rod. Such a model was already considered experimentally \cite{birringer2008magnetic} and numerically \cite{jorgedomingos,alvarez,jdomingos2}. We aim to explore the interplay between the structure formation and the translational and rotational dynamics under the influence of an external field. Several experimental and theoretical studies focused on single-particle dynamics, which has particularly important applications in actuators \cite{coq}, microfluidics \cite{Dhar} and optical traps \cite{shelton}. However, an increasing interest has been devoted to the collective behavior of magnetic particles subject to external fields, since time-dependent fields are an important tool to control self-organization processes. 

Previous experiments and computer simulations reveal that for sufficient strong rotating fields, the spatial symmetry can be broken by the formation of layers in the field plane \cite{Martin1,Martin2,Elsner}. As a consequence of the rotating fields, an inverted dipolar pair interaction with an in-plane attraction and repulsion along the rotation axis \cite{Elsner,Halsey} replaces, averaged over time, the dipolar interaction. Further interesting phenomena occur when the colloidal magnetic particles are exposed to rotating fields in two-dimensional geometry. In this situation, the time-averaged dipolar potential is purely attractive and long-ranged \cite{Jaeger,Klapp}. The fact that particles follow the field synchronously makes possible to explain the resulting self-organized structures from an equilibrium perspective involving the phase behavior of a many-particle system interacting via a time-averaged inverted dipolar interaction \cite{Martin1,Jaeger,smallenburg}.  

Our aim, in this study, is to analyze the phase-behavior of a two-dimensional system consisting of ferromagnetic peapod-like rods subject to rotating fields. According to the extent of the synchronization of the rod particles with the external field, different phases arise from the competition between of the rod-rod interaction, by tunning the dipole misalignment, and the rod-external field interaction, by tuning the field intensity and the rotation (oscillatory) frequency.

The paper is organized as follows: our model system is introduced in Sec. \ref{model}. The numerical results are presented and discussed in Sec. \ref{results}, and our conclusions are given in Sec. \ref{summary}.    


\section{Model}\label{model}
\begin{figure}[h!]
\centering
 \includegraphics[height=6cm]{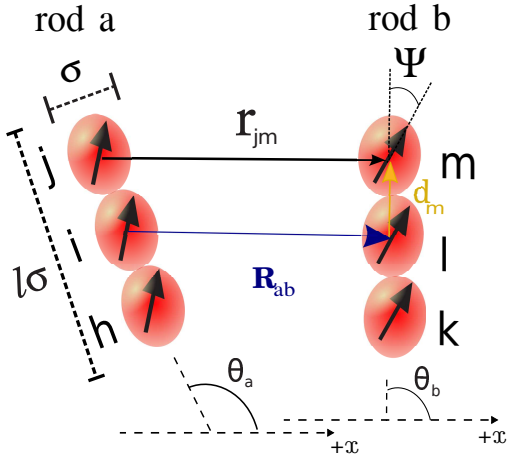}
  \caption{ Schematic illustration of the interaction between two magnetic rods with indication of the important parameters of the pair interaction potential.}
  \label{Fig1}
\end{figure}

We perform extensive Langevin dynamics simulations to study a two-dimensional ($2D$) system consisting of typically $N = 840$ identical stiff rods of aspect ratio $l = 3$. The phase behaviour of a mono-dispersed system with the same aspect ratio was recently studied \cite{may2016colloidal}, being considered a standard reference system \cite{de1991liquid}. For suspensions studied experimentally, the aspect ratio $l = 3$ is in the lowest accessible limit \cite{may2014electric}. We simulate the magnetic nature of the rod by attaching a point dipole of permanent magnetic moment $\mu$ at the center of each bead (see Fig. \ref{Fig1}).

The orientation of the dipoles concerning the axial direction of the rod is given by the angle $\Psi$, as illustrated in Fig \ref{Fig1}.  To model the dipolar particles, we use a dipolar soft sphere (DSS) potential \cite{jorgedomingos}, consisting of the repulsive part of the Lennard-Jones (LJ) potential $u^{rep}$ and a point-like dipole-dipole interaction part $u^{D}$. The total interaction energy between rods $a$ and $b$ is the sum of the pair interaction terms between their respective dipolar spheres (DS):

\begin{eqnarray}
U_{a,b}(\mathbf{R}_{a,b},\theta_a,\theta_b) = \sum_{j \neq m} u_{j,m}\textrm{,} \\
u_{j,m} =  u^{rep}(\mathbf{r}_{jm}^{a,b}) + u^{D}(\mathbf{r}_{jm}^{a,b},\boldsymbol{\mu}_j^{a},\boldsymbol{\mu}_m^{b})\textrm{,}  
\end{eqnarray}
where:

\begin{eqnarray}
u^{rep} &=& 4\epsilon \left(\frac{\sigma}{r_{jm}}\right)^{12}\textrm{,}  \\
u^{D} &=& \frac{\boldsymbol{\mu}_{j}\cdot\boldsymbol{\mu}_{m}}{r_{jm}^3}-
\frac{3(\boldsymbol{\mu}_{j}\cdot\mathbf{r}_{jm})(\boldsymbol{\mu}_{m}\cdot\mathbf{r}_{jm})}{r_{jm}^5}\textrm{,}
\end{eqnarray}

\noindent with $\sigma$ the diameter of each bead, and $\epsilon$ is the LJ soft-repulsion constant, $\mathbf{R}_{a,b} = \mathbf{R}_b - \mathbf{R}_a$ is the vector which connects the center of rod $b$ with the center of rod $a$. The orientation of rods $a$ and $b$ are given by $\theta_a$ and $\theta_b$, respectively. The vector $\textbf{r}_{jm}^{a,b}$ connects the center of bead $m$ of rod $b$ with the center of bead $j$ of rod $a$ (see Fig \ref{Fig1}). The force on bead $m$ due to bead $j$ is given by:

\begin{equation}
\mathbf{f}_{jm} = - \boldsymbol{\nabla} u_{jm}\textrm{.}
\end{equation}

The torque on bead $m$ is \cite{jorgedomingos}:
\begin{eqnarray}\label{torque1}
\boldsymbol{\tau}_{m} = \boldsymbol{\mu}_m \times \sum_{m\neq j}\lbrace\mathbf{B}_{jm}+\mathbf{B}(t)\rbrace + \mathbf{d}_m \times \sum_{m\neq j}\mathbf{f}_{jm}\textrm{,}
\end{eqnarray} 
\noindent where $\textbf{d}_m$ is the vector connecting the center of bead $m$ (rod $b$) with the center of rod $b$ as illustrated in Fig \ref{Fig1}, and $\textbf{B}_{jm}$ is the magnetic field generated by the dipole moment $\mu_j$ at the position of the dipole $\mu_m$, and $B(t)$ is the external magnetic field. They are given by:
\begin{eqnarray}
\mathbf{B}_{jm} = \frac{3(\boldsymbol{\mu}_{m}\cdot\mathbf{r}_{jm})\mathbf{r}_{jm}}{r_{jm}^5}-\frac{\boldsymbol{\mu}_{m}}{r_{jm}^3}\textrm{,}\\
\mathbf{B}(t) = B[cos(\omega t)\mathbf{\hat{x}}+ sin(\omega t)\mathbf{\hat{y}}]\textrm{,}
\end{eqnarray}
\noindent where $B(t)$ is the external magnetic field, with intensity $B$ and rotation frequency $\omega$. The external magnetic field rotates in the same plane of the magnetic rods.

The summations in Eq. \ref{torque1} are considered only for dipoles belonging to distinct rods. The orientation of the rods is given by the unitary vector $\mathbf{s}$ given by $\textbf{s} = \textbf{d}_m/\mid{\textbf{d}_m}\mid$. The translational and rotational Langevin equations of motion of rod $b$ with mass $M_b$ and moment of inertia $I_b$, are given by:
\begin{eqnarray}
M_b\frac{d\mathbf{v}_b}{dt} = \mathbf{F}_b - \boldsymbol{\Gamma_T}\cdot\mathbf{v}_b + \boldsymbol{\xi }^T_b(t)	\textrm{,}
\\
I_b\frac{d\mathbf{\boldsymbol\omega}_b}{dt} = \mathbf{N}_b - \Gamma_R \boldsymbol{\omega}_b + \boldsymbol{\xi }^R_b(t)\textrm{,}
\end{eqnarray}  

\noindent where $\mathbf{v}_b = d\mathbf{R}_b/dt$, $\boldsymbol{\omega }_b$ is the angular velocity,  $\mathbf{F}_b$ and $\mathbf{N}_b$ are the total force and torque acting on rod $b$, respectively, while $\boldsymbol{\Gamma_T}$ and $\Gamma_R$ are the translational tensor and rotational friction parameters. For rod-like particles the translational tensor is composed by the parallel($\zeta_{\parallel}$) and perpendicular ($\zeta_{\perp}$) components with respect to the rod axis, which are given by:
\begin{equation}
\zeta_{\parallel}= \frac{2\pi\eta_0l\sigma}{ln(l) + \delta_{\parallel}}\textrm{,}\quad\zeta_{\perp}= \frac{4\pi\eta_0l\sigma}{ln(l) + \delta_{\perp}}\textrm{,}\label{346}
\end{equation} 
\noindent and for rotation: 
\begin{equation}
\quad\zeta_r =  \frac{\pi \eta_0 ({l\sigma})^3}{3\,ln(l)+\delta_r}\textrm{,}
\end{equation}
\noindent where $\eta_0$ is the solvent viscosity, $\delta_{\parallel}$, $\delta_{\perp}$ and $\delta_r$ are correction factors for small rods extracted from Refs. \cite{tirado1,tirado2}. As a result, the total translational diffusion coefficient is $D_T = \frac{1}{3}(D_{\parallel}+2D_{\perp})$ for $D_{\perp} = \frac{1}{2}D_{\parallel}$ \cite{langrods}.

 $\boldsymbol{\xi }^T_b$ and $ \boldsymbol{\xi }^R_b$ are the Gaussian random force and torque, respectively, which obey the following white noise conditions:$\braket{\boldsymbol{\xi }^\alpha_b(t)} = 0$, $\braket{\boldsymbol{\xi }^\alpha_b(t)\cdot\boldsymbol{\xi }^\alpha_{b'}(t')}=2\Gamma_\alpha k_BT\delta_{bb'}\delta(t-t')$, $\alpha = T,R$.  

  We define the reduced unit of time as $t^* = t/\sqrt{\epsilon^{-1}M\sigma^2}$, where $M$ is the mass of the rod. Therefore, the frequency of the rotation of external magnetic field is $\omega^* = {t^*}^{-1}=\omega/\sigma^2/\sqrt{\epsilon^{-1}M}$. The energy is given in reduced units as $U^* = U/\epsilon$, the dipole moment in dimensionless units as $\mu^{*} = \mu/\sqrt{\epsilon \sigma^3}$, and the dimensionless distances as of $r^* = r/\sigma$. The ratio of the thermal energy to the soft-sphere repulsion constant is chosen to be $k_BT/\epsilon = 1$, where $\epsilon/k_B$ is the temperature unit and $k_B$ is the Boltzmann constant. Periodic boundary conditions are taken in both spatial directions. Since the dipolar pair interaction falls off as $(r^{-3})$, we take the simulation box sufficiently large such that no special long-range summation techniques \cite{allen1989computer} are needed. We define the packing fraction as $\eta = N_{beads}\pi(\sigma/2)^2/L^2$, where $N_{beads} = 2520$ is the total number of dipolar beads of the system and $L^2$ is the simulation box area. Since $N_{beads} = lN$, we can rewrite the packing fraction as $\eta = \rho^*l \pi/4$, where $\rho^*$ is the dimensionless density $\rho^* = \rho\sigma^2$, and $\rho = N/L^2$, in all simulations, we set $\eta = 0.1$. The reduced time step is typically in the range $\delta t^* =  10^{-4} - 10^{-3}$.

The quantities of interest are then averaged over more than $10^6$ time steps. All the beads from all rods have the same dipole moment whose magnitude we set as $\mu^* = 4.4 $ which was estimated based on experiments at room temperature ($T \approx 293 K$) using iron nanoparticles \cite{birringer2008magnetic} with saturation magnetization $M_s(Fe) = 1700$ $kA/m$ and the radius of the particles $r \approx 5$ $nm$. For external magnetic fields, we use $B^*(t)$ values within the range $10\leq B^* \leq 50$, which is related to the experimental range $33$ $mT$ $\leq B \leq 165$ $mT$ at room temperature. Experimental values for the magnetic fields are of the order of $0.1$ $T$ \cite{alvarez}, but ferrofluids have been found to be susceptible already to $B < 10$ $mT$ \cite{birringer2008magnetic}. For the sake of simplification, we are omitting the * superscript hereafter in all dimensionless parameters.

\section{Numerical results}\label{results}
In this section, we present our numerical results. We begin by discussing the conditions for which the different phases are observed for fixed packing fraction, $\eta = 0.1$, and temperature $k_BT/\epsilon = 1$. We study the formation of the clusters according to the parameters $Psi$, $B$, and $\omega$, which are associated to the interaction between rods, and the interaction between the rods and the external magnetic field. We base our analysis on structural and dynamical parameters.

\subsection{Non-Equilibrium Phase Diagram}
In this section we examine the formation of different structures for different values of the magnetic field e rotation frequency. As mentioned previously in this manuscript, there is a relationship between the clustering process and the synchronous rotational motion which is related to the time-averaged dipolar potential in $2D$,
\begin{equation}
u^D(\boldsymbol{r}_{ij}) = \frac{1}{\tau}\int_{0}^{\tau}u(\boldsymbol{r}_{ij},\boldsymbol{\mu}_i(t),\boldsymbol{\mu}_j(t))dt = - \frac{\mu^2}{2r_{ij}^3}\textrm{.}\label{inversedipo} 
\end{equation}
The previous equation is obtained when the magnetic dipoles are in phase, i.e., synchronized with the external rotating magnetic field, resulting in an effective isotropic and attractive pair-interaction potential between the magnetic dipoles. 

\begin{figure}[h!]
\centering
  \includegraphics[height=8cm]{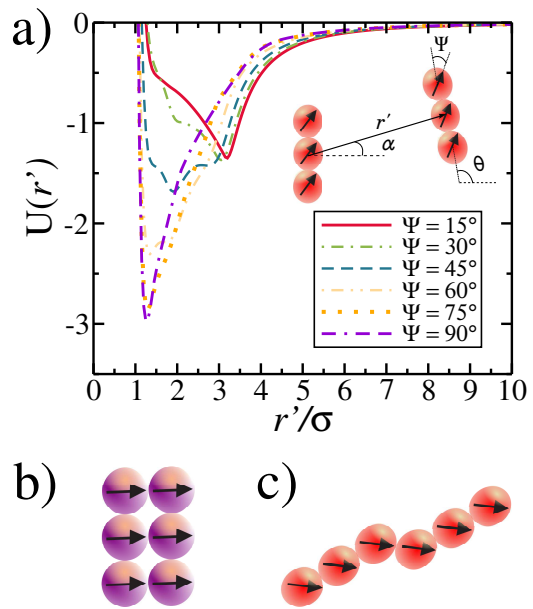}
  \caption{(a) The pair interaction energy as a function of inter-rod separation ($\mathbf{r'}$) minimized with respect to $\boldsymbol{\alpha}$ and $\boldsymbol{\theta}$.  Sketches of the (b) ribbon-like and (c) head-to-tail arrangements.}
  \label{fgr:interação}
\end{figure}

We start by analyzing the dependence of the pair interaction for different values of the new feature added to the rods, the misalignment $\Psi$. We obtain rather different potential profiles by changing $\Psi$. For low values of $\Psi$ ($\leq 30^\circ$), the minima are located at $r'/\sigma \approx 3$, which corresponds to the aspect ratio of the rods. The values of $\alpha$ and $\theta$ [Fig. \ref{fgr:interação}(a)], which minimize the pair-interaction energy indicate that rods are favorably in the head-to-tail bond. By increasing $ \Psi$, the position of the global minimum is displaced to smaller values of $r'/\sigma$, suggesting that the head-to-tail bond disappears, giving rise to the ribbon-like bond configuration (see Fig. \ref{fgr:interação}(b)). The latter arrangement was obtained experimentally in Refs. \cite{tierno,small}.

\begin{figure*}[t!]
\centering
  \includegraphics[height=9.2cm]{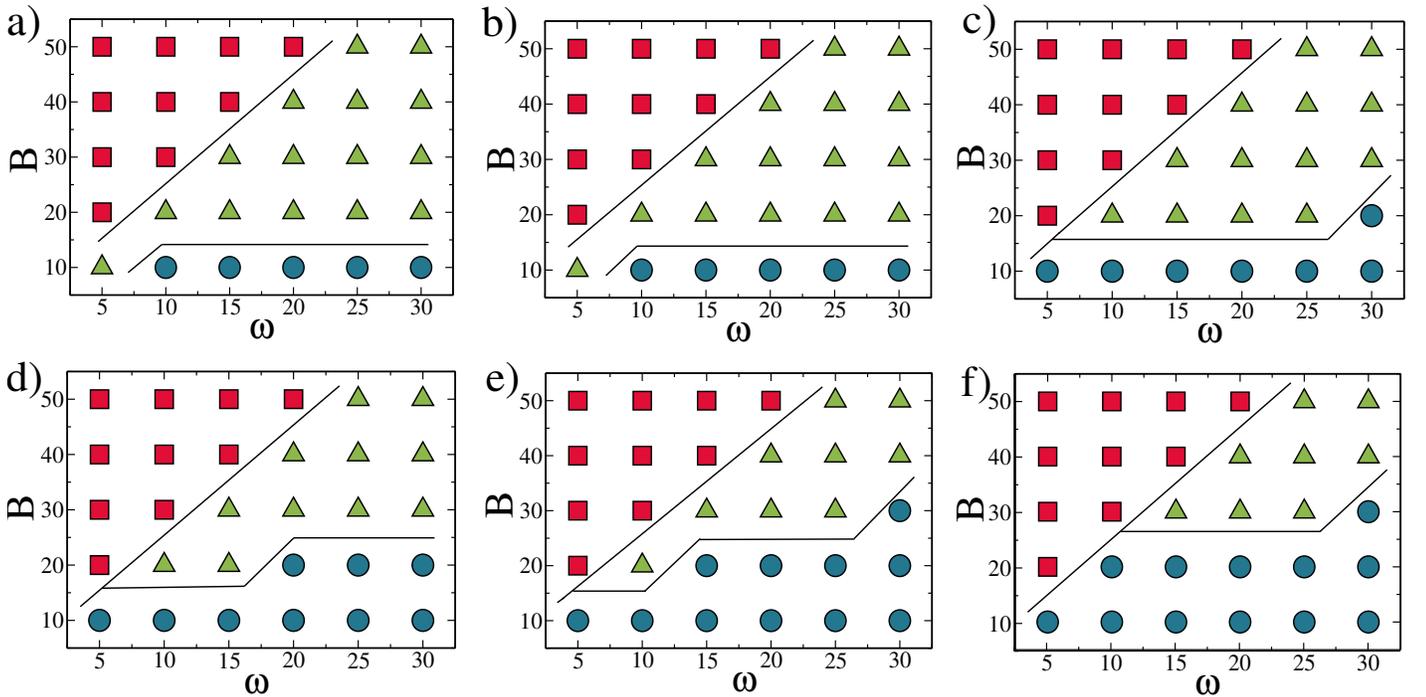}
  \caption{Steady state phase diagrams presenting the self-organized structures as a function of the intensity, $B$, and the rotation frequency, $\omega$, of the external magnetic field for different values of misalignment: (a) $\Psi = 15^\circ$; (b) $\Psi = 30^\circ$; (c) $\Psi = 45^\circ$; d) $\Psi = 60^\circ$; (e) $\Psi = 75^\circ$; (f) $\Psi = 90^\circ$. Symbols represent different phases: $\Box$ Dynamic aggregate, $\bigtriangleup$ Isotropic fluid, $\bigcirc$ Clustered fluid. The solid lines are guides for the eyes to separate regions of different phases.}
  \label{fgr:phasediagram}
\end{figure*}

In order to determine the separation ($\delta_c$) for which we define a bond between two rods, we analyze the inter-rod separation related to the minimum energy value. From Fig. \ref{fgr:interação}(a), the largest inter-rod separation related to the global minimum is located at $ r' \approx 3.4\sigma$ for $\Psi = 15^\circ$ and at $r' \approx 1.4\sigma$ for $\Psi = 90^\circ$. In the former, the rods are in the head-to-tail arrangement, while in the latter they are in the ribbon-like configuration. In both cases, the shortest separation between beads of different rods is $\approx 1.4\sigma$ (bead-to-bead center distance). Therefore, we define that the two rods are bonded as the shortest separation between them is $\leq 1.4\sigma$. 

Since the attraction between rods becomes stronger for larger $\Psi$ (Fig. \ref{fgr:interação}), we expect that the ribbon-like configurations ($\Psi > 45^\circ$) become more stable, implying that the formation of clusters is facilitated in the many-body case. 

The synchronization with the external magnetic field plays, in addition to the dipole's misalignment $\Psi$, an essential role to define the structures fo the system. We observe three distinct typical structures by manipulating $B$ and $\omega$, namely, Dynamic aggregates, Isotropic fluid, and Clustered fluid.  We present in Fig. \ref{fgr:phasediagram} the resulting $B - \omega$ phase diagrams of such configurations for different values of $\Psi$ .

Dynamic aggregates result from a strong interaction between rods and the external magnetic field. If rods follow the rotation of the external field synchronously, an attractive regime appears as a result of the aforementioned time-averaged potential (Eq. (\ref{inversedipo})). In this regime, the rods are not necessarily connected as our definition of a bond suggests (head-to-tail or ribbon-like), and they are rotating with respect to their centers, breaking the spatial symmetry that usually characterizes an isotropic fluid. We show in Fig. \ref{fgr:phasediagram} that the dynamic aggregates appear in the limit of a high magnetic field as $\omega$ increases. These limits allow the majority of the rods to follow the field, fulfilling the behavior of polarizable magnetic particles known to form $2D$ clusters. 

\begin{figure}[h!]
\centering
  \includegraphics[height=3cm]{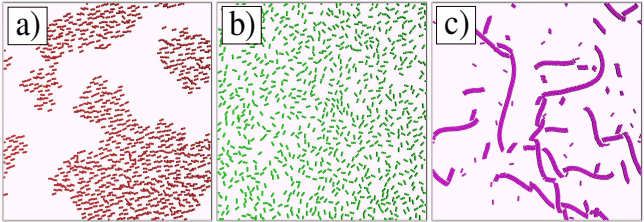}
  \caption{Examples of typical steady state configurations: (a) Dynamical aggregate;(b) Isotropic fluid;(c) Clustered fluid.}
  \label{config1}
\end{figure}

The isotropic fluid is a disordered phase where the kinetic energy is more relevant than the interaction energy between rods. There is no formation of clusters. Instead, we observe a more isotropic particle distribution in the system. For a given magnetic field, the interaction between the rods and the external magnetic field decreases with increasing frequency because of the lack of synchronization, yielding an equivalence of the competition between the rod-rod and rod-external field interactions. The rod-external field interaction is strong enough to avoid the formation clusters, but not strong enough to form dynamic aggregates. When the rod-external field interaction becomes irrelevant, we observe a new phase, the clustered fluid phase, which is a consequence of the rod-rod interaction, and it consists of chains of clustered colloids. In Fig. \ref{config1}, we illustrate examples of the typical phases observed.

Concerning the synchronization of the particles with the external magnetic field,  the dynamical aggregate is in a high synchronization regime (for high $B$ and low $\omega$), while the isotropic fluid phase is in intermediate regime, and the clustered fluid phase is in a low synchronization regime (low $B$ and high $\omega$). We will discuss such regimes in more details in the next section. 

\subsection{Structure and Synchronization}

In this section, we study the structure of the phases and their synchronization with the external magnetic field. Some configurations, even though classified as belonging to the same phase, have distinctive features in their microstructure that make them distinguishable. We base our analysis on the strength of the external magnetic field. One essential tool to analyze the structure of the system is the pair correlation function \cite{rapaport2004art}, defined in $2D$ as:

\begin{equation}
\mathbf{g(r)} = \frac{\Braket{\sum_{a}\sum_{b\neq a}^N\delta(r -R_{ab})}}{2N\pi r \rho^*},
\end{equation}

\noindent where $R_{ab}$ is the separation between centers of rods $a$ and $b$ (see Fig. \ref{Fig1}). To quantify the extent of aggregation between dipolar rods, we analyze the polymerization \cite{polim1}, which is a measure of how many rods are bonded to at least one other rod. We define the polymerization as the ensemble average of the ratio between the number of clustered rods, $N_c$, and the total number of rods, $N$:
\begin{equation}\label{polime}
\Phi = \Braket {\frac{N_{c}}{N}}\textrm{.}
\end{equation}

The synchronization process is based on the analysis of the rotation dynamics of the dipoles by using the single particle time autocorrelation function for the dipole moment (dipole-dipole autocorrelation function), given by:

\begin{equation}\label{eqcorr}
C_{\mu}(t) = \frac{1}{N}\Braket{\sum_{i= 1}^{N}\hat{\mu}_{i}(t)\cdot\hat{\mu}_{i}(0)}\textrm{,}
\end{equation}

\noindent where $\hat{\mu}$ is the unitary vector of the magnetic moment of the $i$-th rod.

\begin{figure}[h!]
\centering
  \includegraphics[height=8cm]{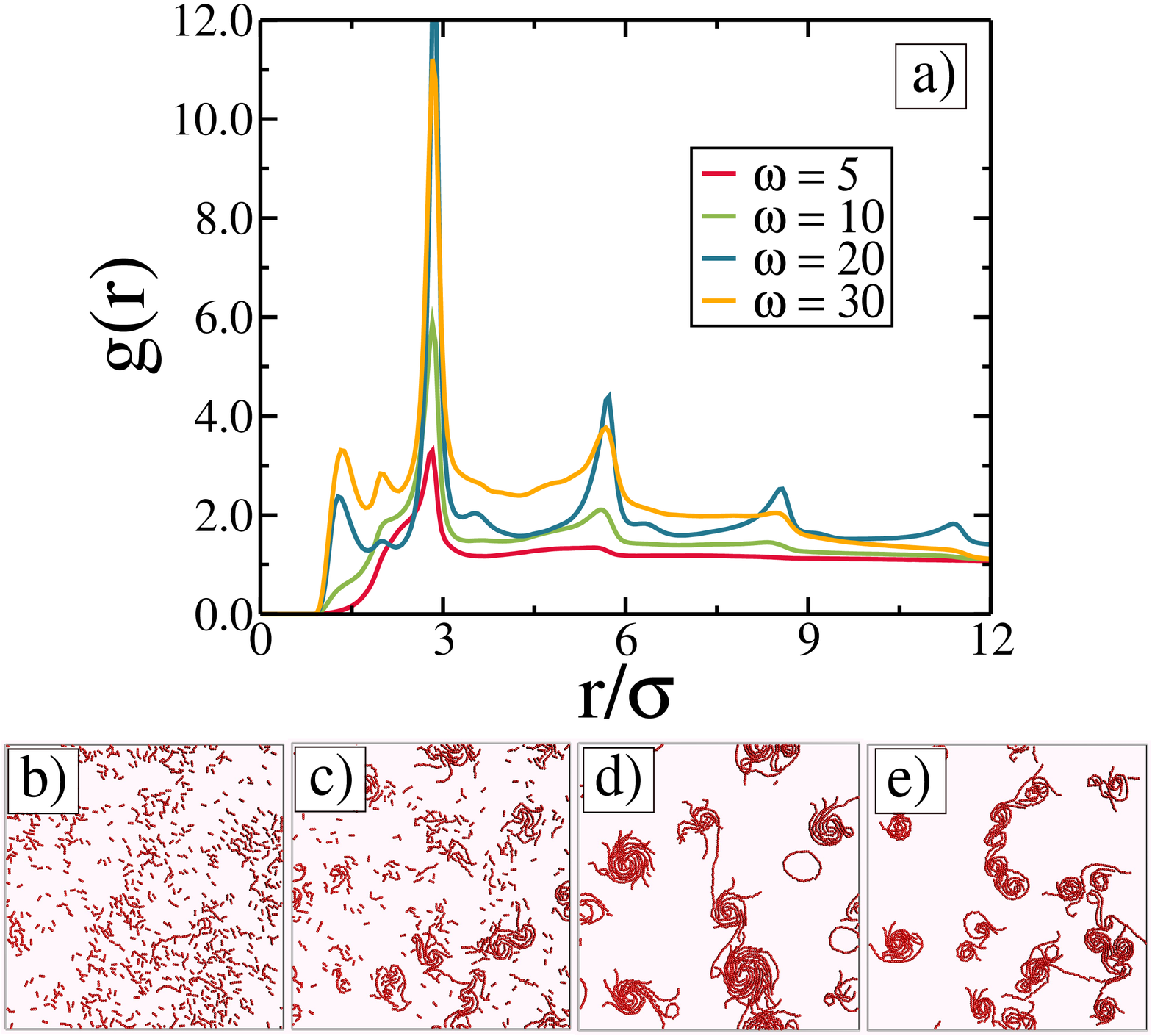}
  \caption{(a) Pair correlation function for $B = 10$ and $\Psi = 15^\circ$ for different values of $\omega$. (b)-(e) Steady state configurations for $\Psi = 15^\circ$ and different rotation frequencies of the external magnetic field: (b) $\omega = 5$; (c) $\omega = 10$; (d) $\omega = 20$; (e) $\omega = 30$.}
  \label{B10Q15} 
\end{figure}

In Fig. \ref{B10Q15} we show the pair correlation function and their respective steady-state structures for $B=10$, $\Psi = 15^\circ$, and different rotation frequencies of the external magnetic field. In the absence of the external magnetic field, the rods tend to cluster together in a head-to-tail arrangement, forming a chain-like structure. As discussed previously, such an arrangement is less stable against fluctuations compared with the ribbon-like configuration. For $B=10$, $\Psi = 15^\circ$, the coupling of the dipole moments with the external field is already strong enough to break the head-to-tail arrangement, especially for low frequency ($\omega = 5$), where some rods tend to follow the external rotating field. We find that $ \approx 20\% $ of the rods are synchronized with the external magnetic field. Those particles avoid the clustering and lead the system to the gas-like configuration shown in Fig. \ref{B10Q15}(a).  For $\omega = 10$, the single-dipole time autocorrelation function, $C_{\mu}(t)$, presents no sine-like oscillation, but a slow decay as a function time, indicating a lack of synchronization between individual rods and the external field, favouring again the formation of the head-to-tail arrangement. As $\omega$ increases, the time-decay of $C_{\mu}(t)$ is even slower, the effective interaction between rods is mostly attractive and strong enough to cluster them together.  For $\omega \geq 10$, the steady states are composed of curled up structures, driven by the external rotating field. Similar structures were observed in a recent study of $3D$ systems of flexible paramagnetic filaments in precessing fields \cite{curl}. For $\omega \ge 20 $ the $g(r)$, in addition to the head-to-tail peaks ($r/\sigma\approx  3$), we also notice some correlation at $r/\sigma< 3$, which is a consequence of the curled up structures.

\begin{figure}[h!]
\centering
  \includegraphics[height=8cm]{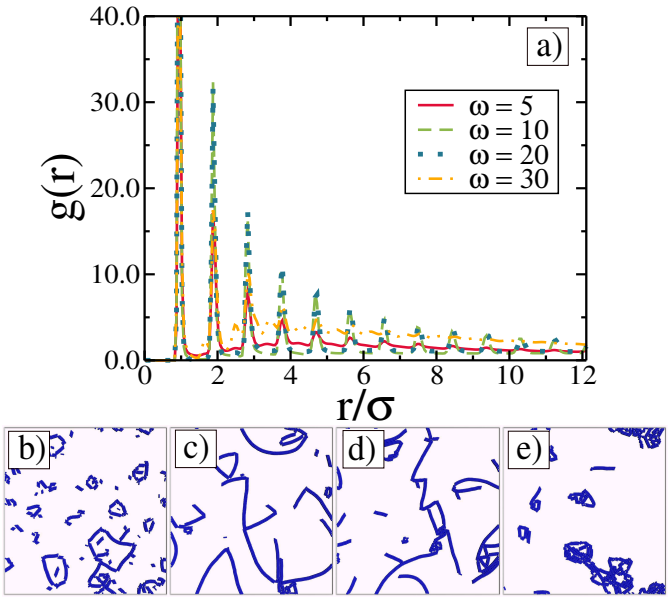}
  \caption{(a) Pair correlation function for $B = 10$ and $\Psi = 45^\circ$ for different values of $\omega$. (b)-(e) Steady states for $\Psi = 45^\circ$ and for different values of $\omega$: (b) $\omega = 5$; (c) $\omega = 10$; (d) $\omega = 20$; (e) $\omega = 30$.}
  \label{B10Q45} 
\end{figure}

\begin{figure}[h!]
\centering
  \includegraphics[height=8cm]{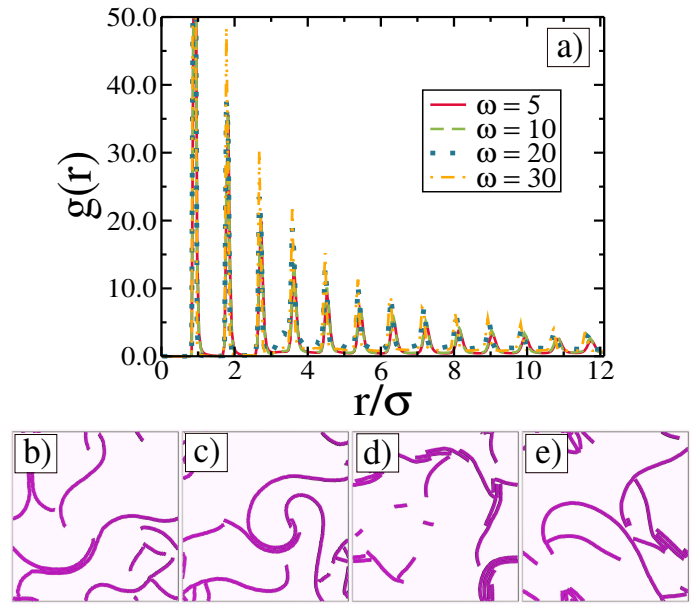}
  \caption{(a) Pair correlation function for $B = 10$ and $\Psi = 90^\circ$ for different values of $\omega$. (b)-(e) Steady states for $\Psi = 90^\circ$ and for different values of $\omega$: (b) $\omega = 5$; (c) $\omega = 10$; (d) $\omega = 20$; (e) $\omega = 30$.}
  \label{B10Q90} 
\end{figure}

For $B=10$, $\Psi = 45^\circ$ there is no synchronization between the dipoles and the external field. The larger deviation of the dipoles from the axial direction of the rods favors the formation of ribbon-like structures, as indicated in the $g(r)$ function presented in Fig. \ref{B10Q45}. Notice that the most intense peaks of $g(r)$ are located at separation $r/\sigma< 3$. The resultant configurations depend strongly on the rotation frequency of the external magnetic field, and we observe an interesting reentrant effect concerning the formation of clusters.  I.e., for sufficiently low ($\omega = 5$) and high ($\omega =30$) rotation frequency of the external magnetic field, cluster-like configurations are  observed, while a linear configuration is found for intermediate values of the rotation frequency ($\omega=10$ and $\omega=20$).  The case $B=10$, $\Psi = 90^\circ$  [Fig. \ref{B10Q90}] corresponds to the strongest rod-rod interaction \cite{jdomingos2}, where the ribbon-like arrangement is dominant, and the clusters are linear and very stable against thermal fluctuations. As indicated in Fig.\ref{B10Q90}, the separation between peaks is $r/\sigma \approx 1$, which confirms the ribbon-like arrangement. There is no important qualitative differences in the configurations as a function of $\omega$. 

\begin{figure}[h!]
\centering
  \includegraphics[height=8cm]{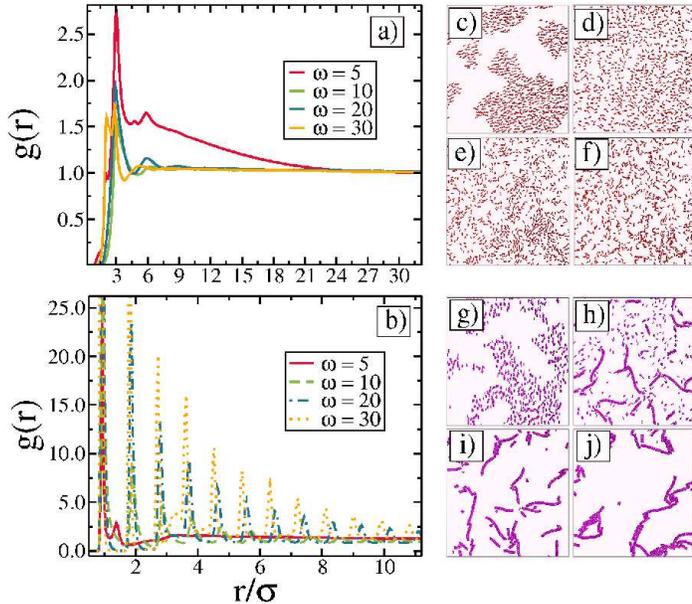}
  \caption{ Pair correlation function for $B = 20$ for different $\omega$, for: (a) $\Psi = 15^\circ$ (b) $\Psi = 90^\circ$. Steady states for $\omega = 5$, $\omega = 10$, $\omega = 20$ and $\omega = 30$, for  $\Psi = 15^\circ$ (c)-(f), and for $\Psi = 90^\circ$ (g)-(j).}
  \label{B20Q1590} 
\end{figure}

We now compare some of the previous results with the ones obtained when the intensity of the external magnetic field is twice larger, $B=20$. In Fig. \ref{B20Q1590} we show the pair correlation function and their respective steady-state structures for $\Psi = 15^\circ$ and $\Psi = 90^\circ$, and different rotation frequencies of the external magnetic field ($B=20$).  For $\Psi = 15^\circ$, the spatial correlation between rods is larger for $\omega=5$ due to the formation of a large cluster, which is favoured by the effective attractive interaction between rods, due to the synchronization of individual rods with the external field.  For higher frequencies, we do not find important differences concerning the range of the spatial correlation, but the micro-structure of the configurations depends on $\omega$  [Fig. \ref{B20Q1590}(a)].  We observe an opposite $\omega$-dependence of the spatial correlation for $\Psi = 90^\circ$, i.e., the range of the spatial correlation increases dramatically for $\omega > 5$, with the rods ordered according to a ribbon-like arrangement [Figs. \ref{B20Q1590}(b),(h),(i),(j)]. Our results of the mean square displacement (not shown in the manuscript) for the cases $\Psi = 15^\circ$ and $\Psi = 90^\circ$  indicate that the system is always in the liquid phase. 

The results presented in Fig. \ref{B20Q1590} can be understood from the coupling between the rods and the external magnetic field. Note that in the limit where the system is fully synchronized (all rods in phase with the external magnetic field), the time-dependence of the orientation of the dipole moment of the rods obeys:
\begin{equation}
\hat{\mu}(t) = cos(\omega t)\hat{x} + sin(\omega t)\hat{y}\textrm{,}
\end{equation}

\noindent so that Eq. (\ref{eqcorr}) results in:
\begin{equation}\label{eq:amplitudecorr}
C_{\mu}(t) = \frac{1}{N}\Braket{\sum_{i= 1}^{N}\hat{\mu}_{i}(t)\cdot\hat{\mu}_{i}(0)} = \frac{n_s}{N}cos(\omega t)\textrm{,}
\end{equation}

\noindent where $n_s$ is the number of synchronized rods ,$N$ is the total number of rods, and $\omega$ is the rotation frequency of the external magnetic field. Therefore, if $C_{\mu}(t)$ presents a sinusoidal oscillatory behaviour,  its amplitude, $n_s/N$, represents the fraction of rods in phase with the external magnetic field.

\begin{figure}[h!]
\centering
  \includegraphics[height=12.cm]{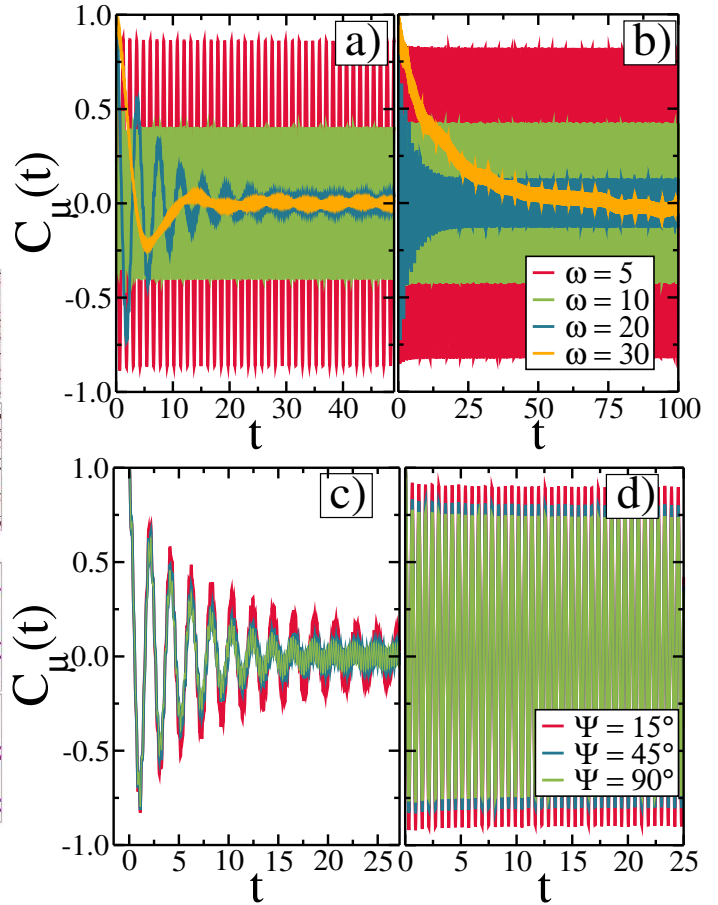}
  \caption{Dipole-dipole autocorrelation function for different cases: (a) $B = 20$ and $\Psi = 15^\circ$;(b) $B = 30$ and $\Psi = 90^\circ$; (c) $B = 30$ and $\omega = 25$; (d) $B = 50$ and $\omega = 10$. The legend in (a) is the same presented in (b). The legend in (c) is the same presented in (d). }
  \label{corB20Y15} 
\end{figure}

For $\Psi=15^\circ$ and $\omega=5$, more than $80\%$  of the rods are in phase with the external field [Fig. \ref{corB20Y15}(a)]. As a consequence, the average interaction between rods is mostly attractive [see Eq. \ref{inversedipo}], leading to the formation of a large polarized cluster (non-zero total magnetic moment) and a larger range of the spatial correlation function, as shown in Fig. \ref{B20Q1590}(a). For $\omega=10$, less than $40\%$ of the rods are in phase with the external field [Fig. \ref{B20Q1590}(a)]. The results indicate that those few synchronized rods avoid the clustering and lead the system to a gas-like configuration. The number of rods in phase with the external field decreases with increasing rotation frequency. For $\omega=30$ such a number is $\lesssim5\%$ [Fig. 
\ref{B20Q1590}(a)]. 

For $\Psi=90^\circ$, we find a partial synchronization only in the case $\omega =5$, where $\sim 70\%$ of the rods are in phase with the external field. A configuration similar to the one of the case ($\Psi=15^\circ$; $\omega=5$) is observed. For $\omega>10$, no synchronization is observed within the simulation time, and due to the perpendicular orientation of the dipoles with respect to the axial axis, the stronger interaction between rods favors the formation of the ribbon-like configuration, which presents a longer spatial order with respect to the previous ones. 

\begin{figure}[h!]
\centering
  \includegraphics[height=6.1cm]{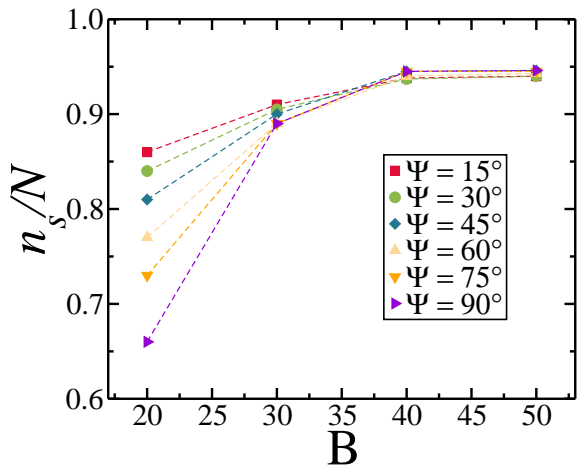}
  \caption{Critical amplitude of oscillation for dynamical aggregate phases for different $\Psi$ as a function of $B$.}
  \label{amptrans} 
\end{figure}

In Fig. \ref{corB20Y15} we also illustrate the behaviour of $C_\mu(t)$ for other values of the parameters $B$, $\omega$ and $\Psi$. By considering the synchronized cases [Figs. \ref{corB20Y15}(a),(b) and (d)] and from Fig. \ref{fgr:phasediagram},  we notice a characteristic amplitude of $C_\mu(t)$, typically associated to the dynamical aggregate phase. In Fig. \ref{amptrans} we show the amplitude of $C_\mu$ [see Eq. \ref{eq:amplitudecorr}] as a function of $B$ for the dynamical aggregate phases considering the $\omega$-interval in which such a phase is found. We notice that the dynamical aggregates are usually observed for $n_s/N \gtrsim 0.65$. 

\begin{figure}[h!]
\centering
\hspace{-0.58pt}\includegraphics[height=5.7cm]{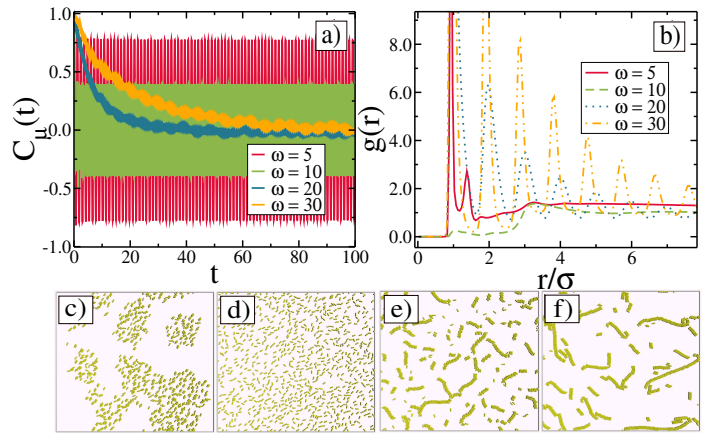}
  \caption{Characterization of $\Psi = 60^\circ$ phases for $B = 20$ for different $\omega$. (a) Dipole-dipole autocorrelation function. (b) Pair correlation function. (c)-(f) Steady states: (c) $\omega = 5$; (d) $\omega = 10$; (e) $\omega = 20$; (f) $\omega = 30$.}
  \label{B20Q60} 
\end{figure}

In Fig. \ref{corB20Y15}(a), the curves for $\omega \geq 10$ are related to the isotropic fluid phase [see Figs. \ref{B20Q1590} (d)-(f)], where $n_s/N \lesssim0.65$ or the autocorrelation function assumes a damped decay with a non-monotonic behaviour with at least a minimum at early times - underdamped behaviour [Fig. \ref{corB20Y15}(a),(b)]. The latter indicates that the reorientation resulting from the rod-rod interaction actuates somehow as a restoring torque, suppressing the rotation of the rods driven by the external field. As a result, it is observed a damped behavior of $C_\mu$. Also, the rotational friction originated from the rod-rod and rod-solvent interactions becomes more important with the decrease of the effect of the external magnetic field. As a result, we observe a damped oscillation behaviour for $C_{\mu}(t)$. These results are in agreement for dipole correlation functions using a rotational diffusion model for a large angle reorientation of the particles \cite{Hansen2}. We also observed the damping decay in Fig. \ref{corB20Y15}(c).
This picture changes either when there is an additional increase of the rod-rod interaction (increase of $\Psi$), or a decrease of the extent of synchronization (increase of $\omega$ or decrease of $B$). The increase of relevance of the rod-rod interaction results in a monotonic average decay of $C_\mu(t)$, as shown e.g. for $\omega = 30$ in Fig. \ref{corB20Y15}(b). We observe that such a time dependence of $C_\mu(t)$ is related to the clustered phase. In Fig. \ref{B20Q60}, we show an example where is possible to observe all phases discussed so far, with their respective related correlations functions. Notice that the pair correlation function (Fig. \ref{B20Q60}(b)) presents for $\omega = 5$ a liquid-like behavior, and a gas-like profile for $\omega =10$, resultant of the dispersed phase, suggesting that we obtain different aggregation states just by tunning the rotation frequency of the external magnetic field.

\begin{table}
\caption{The extent of polymerization $\Phi$ for $B = 20$ as a function of $\omega$ for different $\Psi$. High values at low $\omega$ represent the dynamical aggregate phase}\label{table}
\vspace{0.1cm}
 \includegraphics[height=3.8cm]{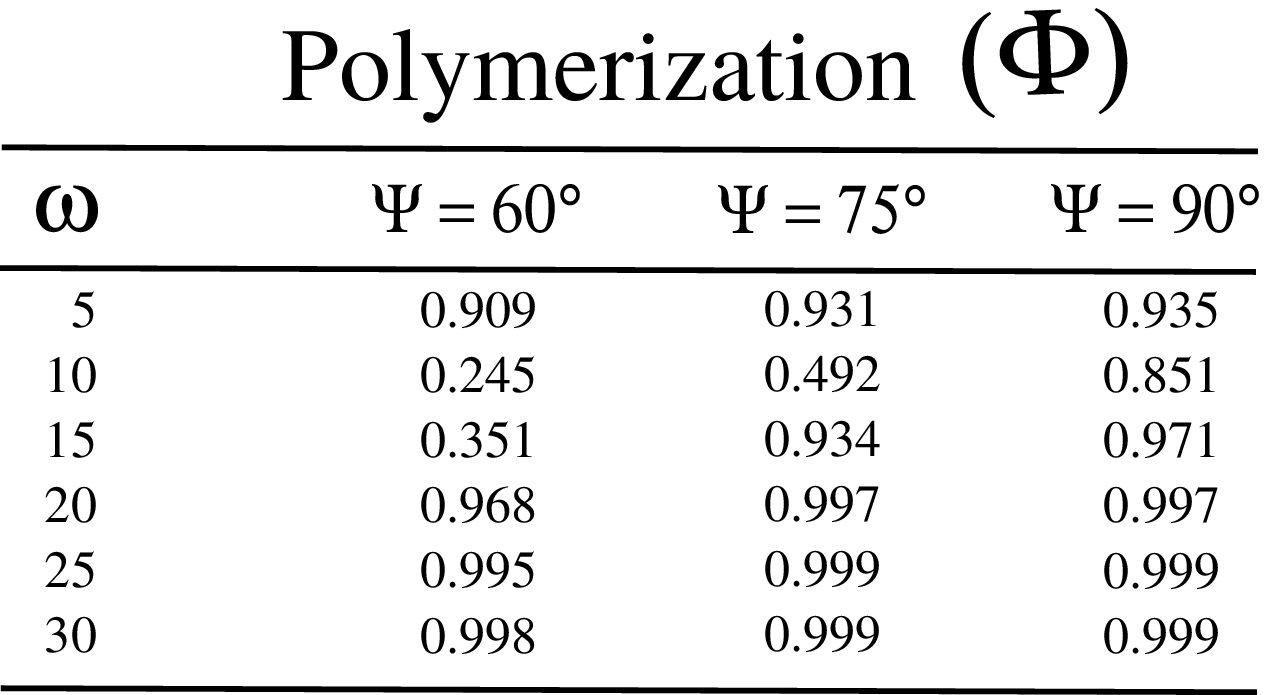}
\end{table}

\begin{figure*}[t!]
\centering
\includegraphics[height=5.8cm]{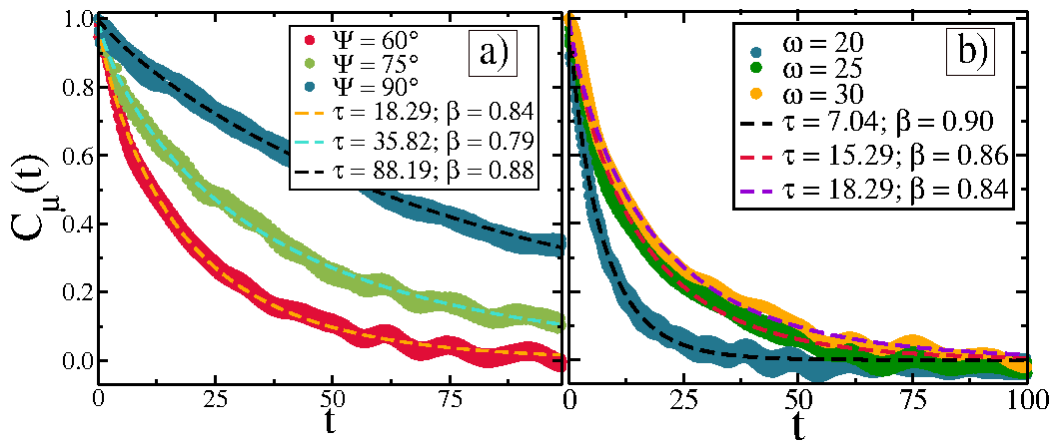}
  \caption{Dipole autocorrelation function for some clustered phases for $B = 20$: (a) for different $\Psi$ and $\omega=30$; (b) for $\Psi = 60^\circ$ and for different $\omega$. The dashed lines represent the stretched exponential fit followed by their respective fit parameters at the legend boxes.}
  \label{fitting}
  \end{figure*}


The clustered fluid phase, typically observed for large $\omega$, presents a polymer-like shape [Fig.  \ref{config1}] and a slower relaxation with increasing degree of polymerization $\Phi$, as shown in Table \ref{table}. Notice that, for the clustered phase, a higher relaxation time is associated with a more polymerized systems.
The decay of the autocorrelation function $C_\mu(t)$ may be fitted with the Kohlrausch-Williams-Watts stretched exponential function, typical of glassy systems \cite{expt}:
\begin{equation}
C_\mu(t) = exp\left[-\left(\frac{t}{\tau}\right)^\beta\right]\textrm{,}
\end{equation}
where $\tau$ is the characteristic relaxation time and $\beta$ is the stretched exponential. 

In the Fig. \ref{fitting}, we present $C_\mu(t)$ for some clustered phases and their respective stretched exponential fits. As discussed previously, the increase of importance of the rod-rod interaction results in an additional friction to rotation, and a slowing-down of the relaxation. Therefore, the characteristic relaxation time increases for stronger rod-rod interaction, and this is due either to the increase of the net interaction itself by increasing $\Psi$ (Fig. \ref{fitting}(a)) or, for a fixed $B$, to the decrease of the effect of the external magnetic field by increasing $\omega$ (Fig. \ref{fitting}(b)). For $\beta = 1$, the exponential decay is characteristic of Debye-like relaxation, however, the behavior of the autocorrelation deviates from the Debye-decay, since we obtained, for all cases , $\beta<1$. Similar non-Debye behaviour was also obtained for a 3D systems of magnetic spherocylinders \cite{alvarez2}. 

\section{\label{summary}Conclusions}
We investigated a self-organization of a two-dimensional system consisting of magnetic peapod rods in the presence of rotating fields using Langevin dynamics simulation.This model was motivated by experimental \cite{birringer2008magnetic} and theoretical \cite{alvarez,jorgedomingos,jdomingos2} studies. Each rod was composed of $3$ soft beads having a central pointlike dipole whose orientation is misaligned with respect the axial axis of the rods. The application of rotating fields in dipolar systems is already known to produce untypical structures, which are a consequence of the resultant time-averaged dipolar interaction \cite{Jaeger,Klapp}. We investigated the configurations as a function of the strength and rotation frequency of the external magnetic field, and the misalignment of the dipoles with respect to the axial direction of the rods.  Particular attention was also addressed to the synchronization resultant of the competition between the rod-rod and rod-external magnetic field interactions. 
 
 We found three different typical steady-state configurations, consequent of the three different synchronization regimes observed. In the high synchronization regime, observed for high strength and low rotation frequency of the external field, we observed the Dynamical aggregates, which are mainly a consequence of the attractive time-averaged dipolar interaction. In the intermediate regime, when the rod-rod and rod-external field interactions are equivalent, the resultant competition produced a dispersed and spatially isotropic fluid. Here, the synchronization with the external magnetic field is not strong enough to produce a dynamical aggregate, but it is sufficient to avoid the rods to form clusters. The third phase is a consequence of the further decrease in synchronization. In this case, the rod-rod interaction is more relevant; as a result, the rods cluster to each other originating the Clustered phase. We also characterized the system by studying the time-dependence of the single dipolar autocorrelation function. We showed that a constant and high amplitude of the autocorrelation oscillation was associated with the high synchronized phase (Dynamical aggregates). A sufficient decrease of that amplitude or a damping behavior of the autocorrelation function was related to the Isotropic fluid phase. The Clustered phase was characterized by a monotonic decay of the autocorrelation function, with a characteristic relaxation time, which increases with increasing of the polymerization of the system. Such a monotonic decay deviated from the Debye relaxation ones.


%



\begin{acknowledgments}
This work was supported by the Brazilian agencies FUNCAP, CAPES and CNPq.
\end{acknowledgments}


\end{document}